\documentclass[aps,prd,preprint,groupedaddress,%
amsmath,amssymb,eqsecnum,showpacs,showkeys]{revtex4}
\usepackage{graphicx}

\providecommand\half{\tfrac{1}{2}}
\providecommand\thrhlf{\tfrac{3}{2}}
\providecommand\thrd{\tfrac{1}{3}}

\DeclareMathOperator{\sech}{sech}

\begin{document}

\title{Dynamics of Tachyonic Dark Matter}

\author{James M.~Starke}

\email{cuastrssa@gmail.com}

\affiliation{The Catholic University of America\\
620 Michigan Avenue NE\\
Washington, D.C.~~20064~~USA}

\author{Ian H.~Redmount}

\email{ian.redmount@slu.edu}

\affiliation{Department of Physics\\
Saint Louis University\\
3511 Laclede Avenue\\
St.~Louis, Missouri~~63103--2010~~USA}

\date{\today}

\begin{abstract}
Usually considered highly speculative, tachyons can be treated via
straightforward Einsteinian dynamics.  Kinetic theory and thermodynamics
for a gas of ``dark'' tachyons are readily constructed.  Such a gas exhibits
density and pressure which, for the dominant constituent of a suitable
Friedmann-Robertson-Walker spacetime, can drive cosmic evolution with
features both similar to and distinct from those of a standard
dark-energy/dark-matter model.  Hence, tachyons might bear further
consideration as a cosmic dark-matter candidate.
\end{abstract}


\maketitle

\section{INTRODUCTION\label{sec1}}

Tachyons---faster-than-light particles~\cite{fein1970}, with spacelike
four-momenta---are usually considered either unphysical or rather fanciful
speculation.  (But see, e.g., Clay and Crouch~\cite{clay1974}.  The
extant literature on tachyons is extensive and diverse.  A more thorough
review than can be accommodated here is in preparation.~\cite{mlr2022})
It is possible, however, to construct straightforward, consistent
Einsteinian dynamics for such particles.  It is further possible
to treat kinetic theory and thermodynamics for a gas of``dark''
(noninteracting) tachyons.  The density and pressure---i.e.,
the equation of state---for such a gas have interesting consequences
if the gas is taken to be the dominant mass/energy constitutent of a
Friedmann-Robertson-Walker cosmological model.  For suitable choices of
parameters, such a model exhibits rapid expansion from an initial singularity,
decelerating to a minimum expansion rate, followed by accelerating expansion.
This is like the behavior of the dark-energy/dark-matter model which has
become the standard model of cosmology, although the late-time behavior
of the two models is different enough that they might be distinguished
by observation or even experiment.  While challenges remain, it may be that
tachyons deserve further investigation as a potential dark-matter candidate.

The mechanics of a single tachyon interacting with simple electric and
magnetic fields---the simplest interaction consistent with Einsteinian
special relativity---is described in Sec.~\ref{sec2}, following.  As
such dynamics undergirds all that follows, these elementary results are
presented in some detail.  The kinetic theory and thermodynamics
of a gas of ``dark'' or noninteracting tachyons is worked out in
Sec.~\ref{sec3}.The consequences of the resulting properties for a
tachyon-dominated homogeneous and isotropic cosmological spacetime are
shown in Sec.~\ref{sec4}.  Conclusions and further questions are discussed
in Sec.~\ref{sec5}.  For conciseness, units with speed of light $c=1$ are
used throughout.  

\section{SINGLE-TACHYON DYNAMICS\label{sec2}}

The simplest equation of motion, consistent with Einsteinian dynamics,
for a single tachyon is the Lorentz force law~\cite{mtw1973}
\begin{subequations}
\begin{equation}
\label{eq01a}
\frac{dp^\alpha}{d\tau}=q{F^\alpha}_\beta\,u^\beta\ ,
\end{equation}
with Faraday field tensor
\begin{equation}
\label{eq01b}
{F^\alpha}_\beta=\begin{pmatrix} 0&E_x&E_y&E_z\\ E_x&0&B_z&-B_y\\
E_y&-B_z&0&B_x\\ E_z&B_y&-B_x&0\\ \end{pmatrix}
\end{equation}
\end{subequations}
Here the tachyon has invariant mass\footnote{``Rest mass'' is a misnomer
for tachyons, as they are never at rest in any inertial reference frame.}
$m=i\mu$---imaginary, because the tachyon has spacelike four-momentum---real
electric charge~$q$, and ``proper time'' $d\tau=i\,d\lambda$, with $\lambda$
a real affine parameter on the spacelike world line of the tachyon.
With four-momentum $p^\alpha=i\mu\,u^\alpha$, this equation can be written
\begin{equation}
\label{eq02}
\frac{dp^\alpha}{d\lambda}=\frac{q}{\mu}\,{F^\alpha}_\beta\,p^\beta\ ,
\end{equation}
entirely in terms of real quantities.  To illustrate the unambiguously
consistent Einsteinian dynamics of tachyons underlying the rest of the results
in this paper, several elementary solutions of this equation are presented
in detail, following.

\subsection{Tachyon in electric field---one dimension\label{sec2a}}

For a tachyon moving in the $x$~direction, with a uniform electric field in
the $x$~direction, equation of motion~\eqref{eq02} reduces to
\begin{subequations}
\begin{align}
\frac{d{\cal E}}{d\lambda}&=\frac{qE_x}{\mu}\,p_x\label{eq03a}\\
\frac{dp_x}{d\lambda}&=\frac{qE_x}{\mu}\,{\cal E}\label{eq03b}
\end{align}
\end{subequations}
with ${\cal E}=p^0$ the energy and $p_x=p^1$ the momentum of the tachyon.
The solution to these, choosing $\lambda=0$ at ${\cal E}=0$ and
$p_x=\mu$, is
\begin{subequations}
\begin{align}
{\cal E}&=\mu\,\sinh(\alpha\lambda)\label{eq04a}\\
p_x&=\mu\,\cosh(\alpha\lambda)\ ,\label{eq04b}
\end{align}
\end{subequations}
with $\alpha\equiv qE_x/\mu$. With
\begin{subequations}
\begin{align}
{\cal E}&=\pm\frac{\mu}{(v^2-1)^{1/2}}\label{eq05a}\\
p_x&=\frac{\mu v}{(v^2-1)^{1/2}}\ ,\label{eq05b}
\end{align}
\end{subequations}
this solution corresponds to
\begin{subequations}
\begin{equation}
v=\coth(\alpha\lambda)\label{eq06a}
\end{equation}
and trajectory
\begin{align}
t(\lambda)&=\frac{1}{\alpha}\,\cosh(\alpha\lambda)\label{eq06b}\\
x(\lambda)&=\frac{1}{\alpha}\,\sinh(\alpha\lambda)\ .\label{eq06c}
\end{align}
\end{subequations}
For $\alpha>0$, this is the upper branch of an hyperbola, describing a tachyon
coming in from $x\to-\infty$ at near light speed with negative energy (moving
backward in time), accelerating to infinite speed and zero energy, then
decelerating to near light speed with increasing positive energy, moving
forward in time and toward $x\to+\infty$.  For $\alpha<0$ this is the lower
branch of the hyperbola, describing a tachyon coming in from $x\to-\infty$ at
near light speed, moving forward in time with positive energy, accelerating
to infinite speed and zero energy, then decelerating to near light speed,
moving toward $x\to+\infty$ with negative energy, backward in time.

The Feinberg~\cite{fein1970} interpretation of these trajectories imposes a
quantum-field-theoretic description of the motion directly on the classical
solution:  The upper branch of the hyperbola describes the production of
a tachyon/antitachyon pair at $t=1/\alpha$ and $x=0$, each at zero energy.
The tachyon moves rightward, decelerating toward light speed, while the
antitachyon moves leftward, also decelerating toward light speed.  Each
particle maintains zero total energy: energy and momentum are both conserved.
The lower branch of the hyperbola describes the corresponding
tachyon/antitachyon annihilation process.

\subsection{Tachyon in electric field---two dimensions\label{sec2b}}

With two electric-field components and two momentum
components, the equation of motion takes the form
\begin{subequations}
\begin{align}
\frac{d{\cal E}}{d\lambda}&=\frac{q}{\mu}\,\mathbf{E\cdot p}\label{eq07a}\\
\frac{dp_x}{d\lambda}&=\frac{q}{\mu}\,E_x\,{\cal E}\label{eq07b}\\
\frac{dp_y}{d\lambda}&=\frac{q}{\mu}\,E_y\,{\cal E}\ .\label{eq07c}
\end{align}
\end{subequations}
These imply
\begin{subequations}
\begin{equation}
\frac{d^2{\cal E}}{d\lambda^2}=\alpha^2{\cal E}\ ,\label{eq08a}
\end{equation}
with, here,
\begin{equation}
\alpha\equiv\frac{q}{\mu}(E_x^2+E_y^2)^{1/2}\ .\label{eq08b}
\end{equation}
\end{subequations}
These have solution
\begin{subequations}
\begin{align}
{\cal E}&={\cal E}_0\,\cosh(\alpha\lambda)
+\frac{q({\bf E\cdot p})_0}{\mu\alpha}\,\sinh(\alpha\lambda)\label{eq09a}\\
p_x&=\frac{q^2({\bf E\cdot p})_0E_x}{\mu^2\alpha^2}\,\cosh(\alpha\lambda)
+\frac{q{\cal E}_0E_x}{\mu\alpha}\,\sinh(\alpha\lambda)\notag\\
&\qquad\qquad+p_x^{(0)}-\frac{q^2({\bf E\cdot p})_0E_x}{\mu^2\alpha^2}
\label{eq09b}\\
p_y&=\frac{q^2({\bf E\cdot p})_0E_y}{\mu^2\alpha^2}\,\cosh(\alpha\lambda)
+\frac{q{\cal E}_0E_y}{\mu\alpha}\,\sinh(\alpha\lambda)\notag\\
&\qquad\qquad+p_y^{(0)}-\frac{q^2({\bf E\cdot p})_0E_y}{\mu^2\alpha^2}\ .
\label{eq09c}
\end{align}
\end{subequations}
This result describes three distinct types of motion.

\subsubsection{Degenerate case:\label{sec2b1}}

The values ${\cal E}_0=0$ and $({\bf E\cdot p})_0=0$ imply
${\cal E}=0$, $p_x=p_x^{(0)}$, and $p_y=p_y^{(0)}$.  That is, the
(infinite-speed) tachyon ``evades'' the electric field.

\subsubsection{One-dimensional case:\label{sec2b2}}

Values ${\cal E}_0=0$, $p_y^{(0)}=0$, and
$E_y=0$ imply that solution~\eqref{eq09a}--\eqref{eq09c}reduces to
solution~\eqref{eq04a}--\eqref{eq04b}, as expected.

\subsubsection{Simplest two-dimensional case:\label{sec2b3}}

Values ${\cal E}_0>0$, $p_y^{(0)}=0$, $E_x=0$, and $E_y>0$ imply
\begin{subequations}
\begin{align}
{\cal E}&={\cal E}_0\,\cosh(\alpha\lambda)\label{eq10a}\\
p_x&=p_x^{(0)}\label{eq10b}\\
p_y&={\cal E}_0\,\sinh(\alpha\lambda)\ .\label{eq10c}
\end{align}
\end{subequations}
These correspond to (ordinary) velocities
\begin{subequations}
\begin{align}
v_x&=v_0\,\sech(\alpha\lambda)\label{eq11a}\\
v_y&=\tanh(\alpha\lambda)\label{eq11b}\\
v&=[1+(v_0^2-1)\,\sech^2(\alpha\lambda)]^{1/2}\ ,\label{eq11c}
\end{align}
\end{subequations}
so the tachyon asymptotically approaches the speed of light from above.
This corresponds to trajectory
\begin{subequations}
\begin{align}
t(\lambda)&=\frac{{\cal E}_0}{\mu\alpha}\,\sinh(\alpha\lambda)\label{eq12a}\\
x(\lambda)&=\frac{{\cal E}_0v_0}{\mu}\,\lambda\label{eq12b}\\
y(\lambda)&=\frac{{\cal E}_0}{\mu\alpha}\,\cosh(\alpha\lambda)\ .
\label{eq12c}
\end{align}
\end{subequations}
Solutions~\eqref{eq04a}--\eqref{eq04b} show that a tachyon pushed in the
direction it is moving slows down, approaching the speed of light from above
with increasing energy.  But solutions~\eqref{eq12a}--\eqref{eq12c} show that
a tachyon pushed perpendicular to its direction of motion does not ``back
up''; it accelerates in the direction it is pushed. 		

\subsection{Tachyon in magnetic field\label{sec2c}}

For a tachyon in a uniform, purely magnetic field, say in the $z$~direction,
equation of motion~\eqref{eq01a} reduces to
\begin{subequations}
\begin{align}
\frac{d{\cal E}}{d\lambda}&=0\label{eq13a}\\
\frac{dp_x}{d\lambda}&=\frac{qB_z}{\mu}\,p_y\label{eq13b}\\
\frac{dp_y}{d\lambda}&=-\frac{qB_z}{\mu}\,p_x\label{eq13c}\\
\frac{dp_z}{d\lambda}&=0\ .\label{eq13d}
\end{align}
\end{subequations}
For suitable choice of the zero of~$\lambda$, this has general solution
\begin{subequations}
\begin{align}
{\cal E}&={\cal E}_0\label{eq14a}\\
p_x&=p_\perp\,\cos(\omega\lambda)\label{eq14b}\\
p_y&=-p_\perp\,\sin(\omega\lambda)\label{eq14c}\\
p_z&=p_z^{(0)}\ ,\label{eq14d}
\end{align}
with $p_\perp$ constant and
\begin{equation}
\omega=\frac{qB_z}{\mu}\ .\label{eq14e}
\end{equation}
\end{subequations}
The tachyon executes uniform helical (or circular, for $p_z^{(0)}=0$) motion
at constant speed about the magnetic field lines.  Again the tachyon
accelerates in the direction it is pushed, perpendicular to its trajectory,
not the opposite direction.

The existence of \textit{electrically charged} tachyons raises several
conundra:  Such a tachyon should emit Cerenkov radiation in vacuum.  This
should be readily detectable~\cite{fein1970}, and the associated loss of
energy should cause the tachyon to accelerate toward infinite speed and
zero energy.  Furthermore, it would be energetically possible for a free
photon to decay into a tachyon/antitachyon pair, if the pair coupled to
the electromagnetic field.  Unless some other principle suppresses this
process, the propagation of photons across cosmic distances should severely
constrain the existence of charged tachyons.  It is possible, however, to
sidestep such questions by considering electrically neutral or ``dark''
tachyons.

\section{KINETIC THEORY AND THERMODYNAMICS OF TACHYON GAS\label{sec3}}

It is possible to determine thermodynamic properties and an equation of state
for a tachyon gas, i.e.,for an ensemble of (neutral, dark) tachyons of mass
$m=i\mu$ and nonnegative energies~${\cal E}$ with probability distribution
\begin{equation}
{\cal P}({\cal E})={\cal N}\,\exp(-\beta{\cal E})\ ,\label{eq15}
\end{equation}
where ${\cal N}$ is a normalization factor to be determined and
$\beta=1/(k_BT)$ is the inverse temperature of the distribution.
This can be done by evaluating the energy-momentum tensor
\begin{equation}
T^{\alpha\beta}=\frac{n}{\mu}\,\langle p^\alpha p^\beta\rangle\ ,\label{eq16}
\end{equation}
with $n$ the number density of the particles in the rest frame of the
ensemble.  The tensor takes the matrix form
\begin{equation}
\label{eq17}
(T^{\alpha\beta})=\frac{n}{\mu}\,\begin{pmatrix}
\langle{\cal E}^2\rangle&0&0&0\\
0&\thrd\langle p^2\rangle&0&0\\
0&0&\thrd\langle p^2\rangle&0\\
0&0&0&\thrd\langle p^2\rangle\\ \end{pmatrix}\ ,
\end{equation}
angled brackets denoting ensemble averages.

\subsection{Kinetic theory\label{sec3a}}

Thermodynamic averages for this tachyon gas can be expressed as a related
sequence of integrals.  The normalization constant~${\cal N}$ for the
energy distribution is determined by the condition
\begin{subequations}
\begin{align}
1&={\cal N}\int_{\mu}^\infty4\pi p^2\,e^{-\beta{\cal E}}\,dp\notag\\
&=4\pi{\cal N}\int_0^\infty({\cal E}^2+\mu^2)^{1/2}{\cal E}\,
e^{-\beta{\cal E}}\,d{\cal E}\notag\\
&=-4\pi{\cal N}\frac{\partial}{\partial\beta}\int_0^\infty
({\cal E}^2+\mu^2)^{1/2}\,e^{-\beta{\cal E}}\,d{\cal E}\ ,\label{eq18a}
\end{align}
where the energy/momentum relation ${\cal E}^2-p^2=-\mu^2$ implies
$p\,dp={\cal E}\,d{\cal E}$.  The average energy is given by
\begin{align}
\langle{\cal E}\rangle&=4\pi{\cal N}\int_0^\infty
({\cal E}^2+\mu^2)^{1/2}{\cal E}^2\,e^{-\beta{\cal E}}\,d{\cal E}\notag\\
&=4\pi{\cal N}\frac{\partial^2}{\partial\beta^2}\int_0^\infty
({\cal E}^2+\mu^2)^{1/2}\,e^{-\beta{\cal E}}\,d{\cal E}\ .\label{eq18b}
\end{align}
The mean squared energy is
\begin{align}
\langle{\cal E}^2\rangle&=4\pi{\cal N}\int_0^\infty
({\cal E}^2+\mu^2)^{1/2}{\cal E}^3\,e^{-\beta{\cal E}}\,d{\cal E}\notag\\
&=-4\pi{\cal N}\frac{\partial^3}{\partial\beta^3}\int_m^\infty
({\cal E}^2+\mu^2)^{1/2}\,e^{-\beta{\cal E}}\,d{\cal E}\ ;\label{eq18c}
\end{align}
\end{subequations}
the mean squared momentum is found from $\langle p^2\rangle=\langle{\cal E}^2
\rangle+\mu^2$.	 These can be expressed in forms
\begin{subequations}
\begin{equation}
\label{eq19a}
\mathcal{N}=\frac{\beta}{2\pi^2\mu^2\left[\mathbf{H}_2(\mu\beta)-N_2(\mu\beta)
-\dfrac{2\mu\beta}{3\pi}\right]}\ ,
\end{equation}
\begin{equation}
\label{eq19b}
\langle\mathcal{E}\rangle=3k_BT\,\frac{\mathbf{H}_2(\mu\beta)-N_2(\mu\beta)
-\dfrac{\mu\beta}{3}\,[\mathbf{H}_1(\mu\beta)-N_1(\mu\beta)]}
{\mathbf{H}_2(\mu\beta)-N_2(\mu\beta)-\dfrac{2\mu\beta}{3\pi}}\ ,
\end{equation}
\begin{equation}
\label{eq19c}
\langle\mathcal{E}^2\rangle=12(k_BT)^2\,\frac{\mathbf{H}_2(\mu\beta)
-N_2(\mu\beta)-\dfrac{\mu\beta}{4}\,[\mathbf{H}_1(\mu\beta)-N_1(\mu\beta)]}
{\mathbf{H}_2(\mu\beta)-N_2(\mu\beta)-\dfrac{2\mu\beta}{3\pi}}-\mu^2\ ,
\end{equation}
and
\begin{equation}
\label{eq19d}
\langle p^2\rangle=12(k_BT)^2\,\frac{\mathbf{H}_2(\mu\beta)
-N_2(\mu\beta)-\dfrac{\mu\beta}{4}\,[\mathbf{H}_1(\mu\beta)-N_1(\mu\beta)]}
{\mathbf{H}_2(\mu\beta)-N_2(\mu\beta)-\dfrac{2\mu\beta}{3\pi}}\ ,
\end{equation}
\end{subequations}
featuring Struve functions~$\mathbf{H}_1$ and~$\mathbf{H}_2$ and Neumann
functions~$N_1$ and~$N_2$.

These results can be compared with those for a gas of
ordinary---slower-than-light or \textit{bradyonic}---particles with real
invariant mass~$m_b$.  In this case the normalization constant is given by
\begin{subequations}
\begin{align}
\label{eq20a}
{\cal N}_b&=\left(-4\pi\frac{\partial}{\partial\beta}\int_{m_b}^\infty
({\cal E}^2-m_b^2)^{1/2}\,e^{-\beta{\cal E}}\,d{\cal E}\right)^{-1}\notag\\
&=\frac{1}{4\pi m_b^3}\frac{m_b\beta}{K_2(m_b\beta)}\ ,
\end{align}
the average energy by
\begin{align}
\label{eq20b}
\langle{\cal E}\rangle_b&=4\pi{\cal N}_b\frac{\partial^2}{\partial\beta^2}
\int_{m_b}^\infty({\cal E}^2-m_b^2)^{1/2}\,e^{-\beta{\cal E}}\,d{\cal E}
\notag\\
&=3k_BT+m_b\frac{K_1(m_b\beta)}{K_2(m_b\beta)}\ ,
\end{align}
the mean squared energy by
\begin{align}
\label{eq20c}
\langle{\cal E}^2\rangle_b&=-4\pi{\cal N}_b\frac{\partial^3}{\partial\beta^3}
\int_{m_b}^\infty({\cal E}^2-m_b^2)^{1/2}\,e^{-\beta{\cal E}}\,d{\cal E}
\notag\\
&=m_b^2+12(k_BT)^2+\frac{3m_bk_BT\,K_1(m_b\beta)}{K_2(m_b\beta)}\ ,
\end{align}
and the mean squared momentum by
\begin{align}
\label{eq20d}
\langle p^2\rangle_b&=\langle{\cal E}^2\rangle_b-m_b^2\notag\\
&=12(k_BT)^2+\frac{3m_bk_BT\,K_1(m_b\beta)}{K_2(m_b\beta)}\ ,
\end{align}
\end{subequations}
here with subscript~$b$ distinguishing these bradyonic quantities,
and featuring modified Bessel functions~$K_1$ and~$K_2$.

Results~\eqref{eq19a}--\eqref{eq19d} and~\eqref{eq20a}--\eqref{eq20d}
are general, applicable at any temperature~$T$.  At low
temperatures---$k_BT\ll\mu$ and
$k_BT\ll m_b$---averages~\eqref{eq19b}--\eqref{eq19d}
take the limiting forms
\begin{subequations}
\begin{equation}
\label{eq21a}
\lim_{k_BT\to0}\langle\mathcal{E}\rangle\sim2k_BT\,\left[1+O\left(
\frac{(k_BT)^2}{\mu^2}\right)\right]\ ,
\end{equation}
\begin{equation}
\label{eq21b}
\lim_{k_BT\to0}\langle\mathcal{E}^2\rangle\sim6(k_BT)^2\,\left[1+O\left(
\frac{(k_BT)^2}{\mu^2}\right)\right]\ ,
\end{equation}
and
\begin{equation}
\label{eq21c}
\lim_{k_BT\to0}\langle p^2\rangle\sim\mu^2+6(k_BT)^2\,\left[1+O\left(
\frac{(k_BT)^2}{\mu^2}\right)\right]
\end{equation}
\end{subequations}
for cold tachyons.  Averages~\eqref{eq20b}--\eqref{eq20d} take the
familiar forms
\begin{subequations}
\begin{equation}
\lim_{k_BT\to0}\langle{\cal E}\rangle_b\sim m_b+\frac{3}{2}k_BT+m_b\,O\left(
\frac{k_BT}{m_b}\right)^2\ ,\label{eq22a}
\end{equation}
\begin{align}
\lim_{k_BT\to0}\langle{\cal E}^2\rangle_b&\sim m_b^2
+2m_b\left(\frac{3}{2}k_BT+\frac{15}{8}\frac{(k_BT)^2}{m_b}\right)\notag\\
&\qquad\qquad+\frac{15}{4}(k_BT)^2+m_b^2O\left(\frac{k_BT}{m_b}\right)^3\ ,
\label{eq22b}
\end{align}
and
\begin{equation}
\lim_{k_BT\to0}\langle p^2\rangle_b\sim
2m_b\left(\frac{3}{2}k_BT+\frac{15}{8}\frac{(k_BT)^2}{m_b}\right)
+\frac{15}{4}(k_BT)^2+m_b^2O\left(\frac{k_BT}{m_b}\right)^3\label{eq22c}
\end{equation}
\end{subequations}
for cold bradyons.  At high temperatures---$k_BT\gg\mu$ and
$k_BT\gg m_b$---the averages approach limits
\begin{subequations}
\begin{equation}
\label{eq23a}
\lim_{k_BT\to\infty}\langle{\cal E}\rangle\sim3k_BT-\half\mu\left\{
\frac{\mu}{k_BT}-\frac{\mu^2}{(k_BT)^2}+O\left[\left(\frac{\mu}{k_BT}
\right)^3\ln\left(\frac{\mu}{k_BT}\right)\right]\right\}\ ,
\end{equation}
\begin{equation}
\label{eq23b}
\lim_{k_BT\to\infty}\langle{\cal E}^2\rangle\sim12(k_BT)^2
-\tfrac{5}{2}\mu^2\left[1+O\left(\frac{\mu}{k_BT}\right)\right]\ ,
\end{equation}
\begin{equation}
\label{eq23c}
\lim_{k_BT\to\infty}\langle p^2\rangle\sim12(k_BT)^2
-\thrhlf\mu^2\left[1+O\left(\frac{\mu}{k_BT}\right)\right]\ ,
\end{equation}
\begin{equation}
\lim_{k_BT\to\infty}\langle{\cal E}\rangle_b\sim3k_BT+m_b\left\{
\frac{m_b}{2k_BT}+O\left[\left(\frac{m_b}{k_BT}\right)^3\ln\left(
\frac{m_b}{k_BT}\right)\right]\right\}\ ,\label{eq23d}
\end{equation}
\begin{equation}
\lim_{k_BT\to\infty}\langle{\cal E}^2\rangle_b\sim12(k_BT)^2\left[
1+O\left(\frac{m_b}{k_BT}\right)^2\right]\ ,\label{eq23e}
\end{equation}
and
\begin{equation}
\lim_{k_BT\to\infty}\langle p^2\rangle_b\sim12(k_BT)^2\left[
1+O\left(\frac{m_b}{k_BT}\right)^2\right]\ .\label{eq23f}
\end{equation}
\end{subequations}
The behaviors of both hot tachyons and hot bradyons approach that of photons.

\subsection{Equations of state\label{sec3b}}

Energy-momentum tensor~\eqref{eq17} implies tachyon-gas equation of state
\begin{equation}
P=\frac{1}{3}(\rho+\mu n)\ ,\label{eq24}
\end{equation}
with $\rho$ the energy density and $n$ the particle-number density of the
gas.  The corresponding bradyon-gas equation of state is
\begin{equation}
P_b=\frac{1}{3}(\rho_b-m_bn_b)\ ,\label{eq25}
\end{equation}
differing only in the sign of the mass term.  The low-temperature limit
of Eq.~\eqref{eq24} is
\begin{equation}
\label{eq26}
\rho\sim18P\,\left(\frac{k_BT}{\mu}\right)^2\ ;
\end{equation}
the pressure approaches finite limit~$\thrd\mu n$, while the energy density
approaches zero.  In contrast, the low-temperature limit of
Eq.~\eqref{eq25} is
\begin{equation}
P_b\sim\rho_b\frac{k_BT}{m_b}\ ,\label{eq27}
\end{equation}
the familiar Ideal Gas Law.  The high-temperature limits of the equations
are $P\sim\thrd\rho$ and $P_b\sim\thrd\rho_b$; again, both tachyons and
bradyons behave like photons at high temperatures.

\section{TACHYON-DOMINATED COSMOLOGY\label{sec4}}

The equation of state determines the behavior of the gas as a constituent
of an homogeneous, isotropic spacetime.  The Einstein Field Equations,
applied to a Friedmann-Robertson-Walker (FRW) spacetime with scale factor
(curvature radius)~$a(t)$ and comoving-observer time~$t$, can be cast
in the forms
\begin{subequations}
\begin{equation}
\label{eq28a}
3\,\frac{\left(\dfrac{da}{dt}\right)^2+k}{a^2}=8\pi G\,\rho_{\rm tot}
\end{equation}
and
\begin{equation}
\label{eq28b}
\frac{d}{dt}(\rho_{\rm tot}\,a^2)+p_{\rm tot}\,\frac{d}{dt}a^3=0\ .
\end{equation}
\end{subequations}
Here $\rho_{\rm tot}$ is the total energy density and $p_{\rm tot}$ the total
pressure of the cosmic fluid, $k\in\{-1,0,+1\}$ is the curvature parameter
corresponding to an open (hyperbolic), spatially flat, or closed universe,
respectfully, and $G$ is the Newton gravitational constant.

The second of these equations is essentially the First Law of Thermodynamics
for a comoving volume of fluid.  Bradyonic equation of state~\eqref{eq25}
applied to this equation implies density
\begin{equation}
\rho_b(a)=\frac{(\rho_{b0}-{\cal M}_{b0})\,a_0^4}{a^4}
+\frac{{\cal M}_{b0}\,a_0^3}{a^3}\ ,\label{eq29}
\end{equation}
where ${\cal M}_b=m_bn_b$ denotes rest-mass density of the gas and
subscript~$0$ identifies values at some chosen fiducial time, e.g.,
the present time.  This is exactly as expected:  The rest-mass density
undergoes dilution as the spacetime expands, and varies as~$1/a^3$;
the kinetic-energy density undergoes both dilution and redshift,
varying as~$1/a^4$.  As a source for Friedmann equation~\eqref{eq28a},
this engenders the same behavior as a ``traditional'' FRW model with matter
and radiation constituents.  (What is now called the ``standard model'' of
cosmology also includes a ``dark energy'' term corresponding to a different
equation of state, e.g., a vacuum-energy or cosmological-constant term with
equation of state $p_\Lambda=-\rho_\Lambda$.)

\begin{figure}
\includegraphics{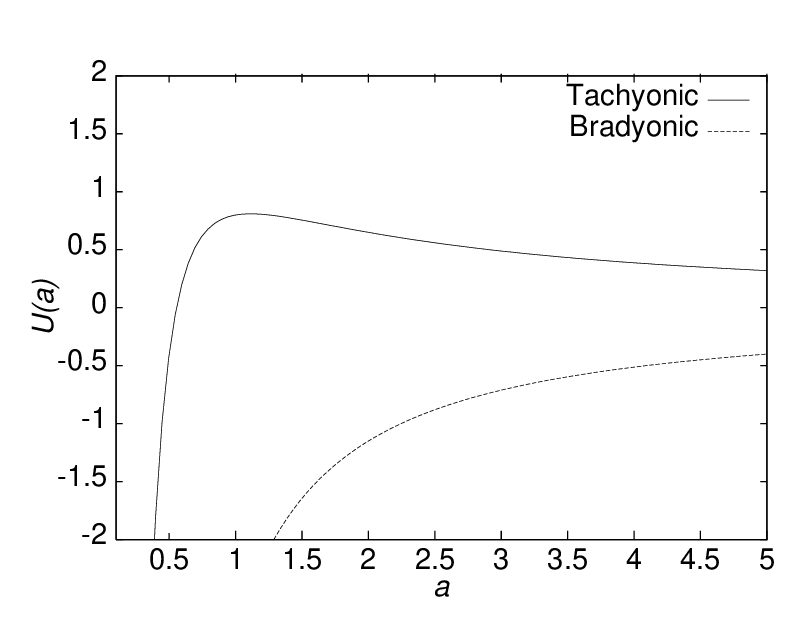}
\caption{\label{f01}Friedmann-equation ``potential energy'' $U(a)$ for
tachyonic (solid curve) and bradyonic (dashed curve) gases.  The tachyonic
curve corresponds to Eq.~\eqref{eq32a}, with $A=1.0$ and $B=0.9$, in the
same arbitrary units as scale factor~$a$.  The bradyonic curve corresponds
to the same parameter values, the $B$ term taking a negative sign.
Closed, spatially flat, and open FRW models correspond to constant
total ``energies''~-1, 0, and~+1, respectively, on this figure.} 
\end{figure}

But tachyonic equation of state~\eqref{eq24} combined with Eq.~\eqref{eq28b}
implies density
\begin{equation}
\rho(a)=\frac{(\rho_0+{\cal M}_0)\,a_0^4}{a^4}
-\frac{{\cal M}_0\,a_0^3}{a^3}\ ;\label{eq30}
\end{equation}
the invariant-mass density terms appear with the opposite sign.  This
difference is more significant than it may appear.  This density will violate
the Weak Energy Condition---become negative---for sufficiently large values
of the scale factor~$a$; this must be regarded as an analytic continuation
of form~\eqref{eq17}.  The Friedmann equation for a model dominated by this
tachyonic density can be cast in the form
\begin{equation}
\label{eq31}
\left(\frac{da}{dt}\right)^2-\frac{8\pi G}{3}\left(
\frac{(\rho_0+{\cal M}_0)\,a_0^4}{a^2}-\frac{{\cal M}_0a_0^3}{a}\right)
=-k\ .
\end{equation}
This can actually be interpreted in \textit{exactly} the same way as the
energy equation used to determine the speed of a car on a roller coaster,
neglecting friction and drag:  The terms on the left correspond to kinetic
and potential energy, respectively; that on the right is the conserved
total energy.  The behavior of the ``potential energy'' term is illustrated
in Fig.~\ref{f01}.
It can be parametrized as
\begin{subequations}
\begin{equation}
U(a)=\frac{2B}{a}-\frac{A^2}{a^2}\ ,\label{eq32a}
\end{equation}
with
\begin{equation}
A\equiv\left(\frac{8\pi G(\rho_0+{\cal M}_0)a_0^4}{3}\right)^{1/2}
\label{eq32b}
\end{equation}
and
\begin{equation}
B\equiv\frac{4\pi G{\cal M}_0a_0^3}{3}\ ,\label{eq32c}
\end{equation}
\end{subequations}
both parameters with units of length.  The tachyonic potential diverges
to~$-\infty$ as $a\to0^+$, crosses from negative to positive values at
$a=A^2/(2B)$, reaches a peak at $a=A^2/B$ of height $B^2/(A^2)$, then
decreases asymptotically to zero as $a\to+\infty$.  Hence, all closed
($k=+1$) and spatially flat ($k=0$) tachyon-dominated FRW spacetimes
must expand from zero scale factor, reach a maximum scale, and
recollapse symmetrically.  The behavior of open ($k=-1$) models depends
on the values of~$A$ and~$B$, i.e., on the energy and mass densities of
the tachyon gas.  For $B>A$,  the solutions must either expand from $a=0$
to a maximum and recollapse, or collapse from $a\to+\infty$ to a minimum
and reexpand.  For $B=A$, there are five possibilities:  The spacetime
can be static at $a=A=B$; or it can asymptotically approach this value
from $a=0$; or it can collapse to $a=0$ from this value, departing arbitarily
slowly from the initial value; or it can collapse from $a\to+\infty$
asymptotically to $a=A=B$; or it can expand to $a\to+\infty$ from this
value, again departing arbitrarily slowly from the initial value.  For $B<A$,
the solutions expand from $a=0$; the expansion slows near $a=A^2/B$, then
accelerates.  \textit{This is precisely the behavior which astronomical
observations demand of the ``standard model'' of cosmology.}  These
open tachyon-dominated  models expand to $a\to+\infty$, the rate of
expansion asymptotically approaching unity, i.e.,~$c$.

All of these behaviors are exhibited in exact solutions of the Friedmann
Equation~\eqref{eq31}.  In terms of conformal time~$\eta$ defined via
$d\eta=dt/a(t)$, closed models with $k=+1$ satisfy
\begin{subequations}
\begin{align}
a(\eta)&=A\,\sin\eta-B(1-\cos\eta)\label{eq33a)}\\
t(\eta)&=A(1-\cos\eta)-B(\eta-\sin\eta)\ ,\label{eq33b}
\end{align}
\end{subequations}
for $0\le\eta\le2\tan^{-1}(A/B)$.  Spatially flat $k=0$ models satisfy
\begin{subequations}
\begin{align}
a(\eta)&=A\eta-\half B\eta^2\label{eq34a}\\
t(\eta)&=\half A\eta^2-\tfrac{1}{6} B\eta^3\label{eq34b}
\end{align}
\end{subequations}
for $0\le\eta\le2A/B$.  Open $k=-1$ models with $B>A$ are described
either by
\begin{subequations}
\begin{align}
a(\eta)&=A\,\sinh\eta-B(\cosh\eta-1)\label{eq35a}\\
t(\eta)&=A(\cosh\eta-1)-B(\sinh\eta-\eta)\label{eq35b}
\end{align}
\end{subequations}
for $0\le\eta\le2\,\tanh^{-1}(A/B)$, or
\begin{subequations}
\begin{align}
a(\eta)&=B+(B^2-A^2)^{1/2}\cosh\eta\label{eq36a}\\
t(\eta)&=B\eta+(B^2-A^2)^{1/2}\sinh\eta\label{eq36b}
\end{align}
\end{subequations}
for $\eta\in(-\infty,+\infty)$. For $B=A$ the five possibilites are:
\begin{subequations}
\begin{align}
a(\eta)&=B(1-e^{-\eta})\label{eq37a}\\
t(\eta)&=B[\eta-(1-e^{-\eta})]\label{eq37b}
\end{align}
for $\eta\in[0,+\infty)$, or its time reversal
\begin{align}
a(\eta)&=B(1-e^{\eta})\label{eq37c}\\
t(\eta)&=B[\eta+(1-e^{\eta})]\label{eq37d}
\end{align}
\end{subequations}
for $\eta\in(-\infty,0]$; the static solution
\begin{subequations}
\begin{align}
a(\eta)&=B\label{eq38a}\\
t(\eta)&=B\,\eta\ ;\label{eq38b)}
\end{align}
\end{subequations}
or
\begin{subequations}
\begin{align}
a(\eta)&=B(1+e^\eta)\label{eq39a}\\
t(\eta)&=B(\eta+e^\eta)\label{eq39b)}
\end{align}
for $\eta\in(-\infty,+\infty)$, or its time reversal
\begin{align}
a(\eta)&=B(1+e^{-\eta})\label{eq39c}\\
t(\eta)&=B(\eta-e^{-\eta})\label{eq39d}
\end{align}
\end{subequations}
for $\eta\in(-\infty,+\infty)$.  For $B<A$ the solution again takes the form
\begin{subequations}
\begin{align}
a(\eta)&=A\,\sinh\eta-B(\cosh\eta-1)\label{eq40a}\\
t(\eta)&=A(\cosh\eta-1)-B(\sinh\eta-\eta)\ ,\label{eq40b}
\end{align}
\end{subequations}
for $\eta\in[0,+\infty)$ in this case.  It is this last model which most
closely mirrors key features of the standard model.

\begin{figure}
\includegraphics{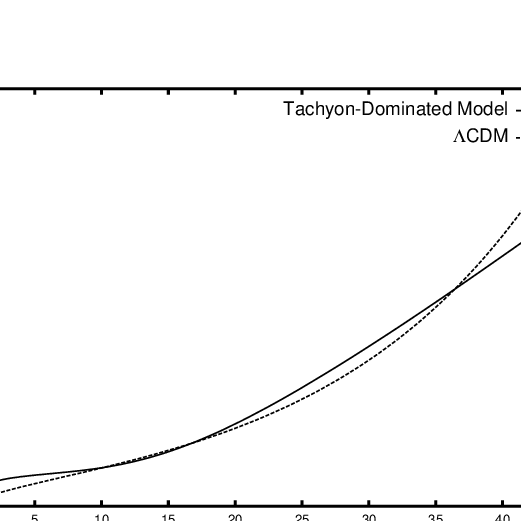}
\caption{\label{f02}Scale factors~$a(t)$ for an open tachyon-dominated
model (solid) and a spatially flat dark-energy-dominated $\Lambda$CDM model
(dashed).  Model parameters are as given in the text.}
\end{figure}

Solution~\eqref{eq40a}--\eqref{eq40b} can readily be compared with a
``standard'' $\Lambda$CDM (cosmological-constant, i.e.,
dark-energy/cold-dark-matter) model.  For a spatially flat ($k-0$) model
with vacuum energy density~$\rho_V$, radiation density~$\rho_r$, and matter
density $\rho_m$, Eq.~\eqref{eq28a} can be separated thus:
\begin{subequations}
\begin{equation}
\label{eq41a}
dt=\dfrac{a\,da}{H_0\,\left(\Omega_{V0}\,a^4+\Omega_{m0}a_0^3\,a
+\Omega_{r0}a_0^4\right)^{1/2}}\ ,
\end{equation}
with Hubble parameter $H\equiv(1/a)(da/dt)$, density ratios
$\Omega_V\equiv\rho_V/\rho_{\rm tot}$, etc., and subscript~$0$ again
denoting present-time values.  This can be integrated:
\begin{equation}
\label{eq41b}
t=H_0^{-1}\,\int_0^{a/a_0}\dfrac{x\,dx}{\left(\Omega_{V0}\,x^4
+\Omega_{m0}\,x+\Omega_{r0}\right)^{1/2}}
\end{equation}
\end{subequations}
to display the behavior~$a(t)$ of the model.  In Fig.~\ref{f02}
an open tachyon-dominated model given by Eqs.~\eqref{eq40a}--\eqref{eq40b},
with parameter values $A=0.930~\hbox{Gpc}$ and $B=0.924~\hbox{Gpc}$
(hence, $a/c=3.03~\hbox{Gyr}$ and $b/c=3.01~\hbox{Gyr}$),
is compared with a spatially flat $\Lambda$CDM model with parameters
$H_0^{-1}=1.37~\hbox{Gyr}$, $a_0=1.40~\hbox{Gpc}$, and density ratios
$\Omega_{V0}=0.7000$, $\Omega_{m0}=0.2998$, and
$\Omega_{r0}=0.0002$.  The parameter values are chosen to give the models
``typical'' features.
Each model exhibits the $a(t)\sim t^{1/2}$ behavior of radiation-dominated models
at early times, and each displays the inflection indicating a change from decelerated
to accelerated expansion, the so-called ``cosmic jerk.''  At late times the
$\Lambda$CDM shows the exponential expansion associated with vacuum-energy
domination; this model approaches de~Sitter space asymptotically.  In constrast,
the tachyon-dominated model shows the linear behavior $a(t)\sim t$ at late
times; the model approaches the Milne~\cite{miln1932} version of flat spacetime.
\section{CONCLUSIONS\label{sec5}}

It is possible to construct straightforward Einsteinian dynamics and
thermodynamics for tachyons.  Tachyons subject to electromagnetic forces
behave exactly as expected, although the existence of charged tachyons
may raise several conundra.  Kinetic theory for a gas of ``dark'' tachyons
implies an equation of state similar in form to that for a gas of bradyons,
save that the invariant-mass density term appears with positive rather than
negative sign, corresponding to much greater pressure in a tachyon gas than
in a bradyon gas of the same energy density and temperature.

For a constituent of a Friedmann-Roberston-Walker spacetime, the tachyonic
equation of state implies a density which varies with scale factor in a manner
similar to that of a bradyon gas, but again with invariant-mass density
contributions of the opposite sign.  As a dominant source in such a spacetime,
this density can give rise to cosmic evolution similar---but not
identical---to that in models containing ordinary bradyonic matter and
radiation.  An open (hyperbolic) tachyon-dominated model can even evolve
with decelerating, then accelerating expansion, similar---but not(
identical---to that of the ``standard model'' containing bradyonic
matter, radiation, and ``dark energy.''  Whether a tachyon-dominated
cosmological model is competitive with the standard model depends on
whether detailed features of the tachyonic model can be matched to the
extensive array of astronomical observations on cosmological scales,
including supernova distances, the cosmic microwave background, and
large-scale structure.  Such a project is the subject of several ongoing
works~\cite{mart2019}, \cite{kram2022}, \cite{gopr2022}.

As noted, the evolution of a tachyon-dominated model can be similar but not
identical to that of the standard model:  The latter continues to accelerate
exponentially in time, while the former asymptotically approaches expansion
at a constant rate.  This difference might even affect the dynamics of
elementary-particle fields~\cite{redm1999}, making it possible to test
the two models via experiments at the scale of such particles.

Of course, the actual existence of tachyons would raise many other questions.
A proper, quantum-field-theoretic description of tachyons would involve a
number of subtleties, such as tachyon ``spin,'' statistics, and interactions
with other quantum fields.  But such a possibility might not be quite as
far fetched as it is usually regarded.




\end{document}